\documentstyle[prl,aps,psfig,epsfig,floats,twocolumn]{revtex} 
\begin{document} 
\def\be{\begin{equation}} 
\def\ee{\end{equation}} 
\def\bc{\begin{center}} 
\def\ec{\end{center}} 
\def\bea{\begin{eqnarray}} 
\def\eea{\end{eqnarray}} 
\draft 
\twocolumn[\hsize\textwidth\columnwidth\hsize\csname 
@twocolumnfalse\endcsname 
\title{Renormalization group  study of
one-dimensional systems with roughening transitions} 
 
\author{G. Bianconi$^{(1)}$, M. A.  Mu{\~{n}}oz $^{(1,2)}$, 
A. Gabrielli$^{(1,3,4)}$, and L. Pietronero$^{(1)}$} 
\address{$^{(1)}$ 
Dipartimento di Fisica and INFM Unit, 
University of Rome ``La Sapienza'', I-00185 Roma} 
\address{$^{(2)}$ The Abdus Salam International 
Centre for Theoretical Physics (ICTP) 
P.O. Box 586, 34100 Trieste, Italy} 
\address{$^{(3)}$  
University of Rome, ``Tor Vergata'', I-00133 Roma} 
\address{$^{(4)}$ 
PMMH-ESPCI,
10, rue Vauquelin, 75231 Paris cedex 05, France} 
\maketitle 
\begin{abstract} 
A recently introduced real space renormalization group technique, 
developed for the analysis of processes in the Kardar-Parisi-Zhang 
universality class, is generalized and tested 
 by applying it to a  
different family of surface growth processes.
 In particular, we consider a
growth model exhibiting a rich phenomenology even in 
one dimension. It has four different phases and a directed
percolation related roughening  
transition. The renormalization 
method reproduces extremely well all the phase diagram, the 
roughness exponents in all the phases and the separatrix  
among them. This proves the versatility of the method and
elucidates interesting physical mechanisms.
\end{abstract} 
\narrowtext 
\vskip2pc] 
{PACS numbers: 05.20.-y, 05.40+j, 05.70.Fh} 
 
\section{Introduction} 

 Surfaces and interfaces may grow in a smooth way or alternatively
in a rough fashion. The study of the physical mechanisms
originating these different behaviors has been the focus of an
overwhelming number of recent studies \cite{HZ,Krug,Laszlo}.
 The Kardar-Parisi-Zhang (KPZ) \cite{KPZ}
equation is the minimal continuous model capturing the aforementioned
physics. In dimensions larger than two it may exhibit two
different phases: a flat and a rough one. Separating both
of them there is a  {\it  roughening transition}.
Apart of being a milestone in surface growth theory, the KPZ equation
is also related to other interesting physical problems:
The Burgers equation in turbulence \cite{Burgers},
 directed polymers in random media \cite{DPRM}, and systems
with multiplicative noise \cite{MN} among others.

 While
the physics of the KPZ flat phase is very well understood,
elucidating
the properties of the rough phase has proven a puzzling problem
 \cite{HZ,Terry,LL}.
 In fact, standard field theoretical analysis finds an unavoidable
difficulty: The rough phase regime is controlled by a
strong coupling fixed point, not accessible by standard
perturbation techniques. Therefore, from  a field theory
 point of view not
much can be concluded about the rough phase  (recent interesting
results in this direction can be found in \cite{LL,FTJ,MB}).
  An alternative strategy has been recently proposed to deal
with this elusive problem; namely a real space renormalization
group (RSRG) approach. Its non-perturbative nature permits a
direct access to the strong coupling regime and, in particular
 gives 
estimations of the roughness exponent in dimensions ranging from
$d=1$ to $d=9$ \cite{First,Second,Third} (see also \cite{Granada}
 with results in very good agreement with the
best numerical measurements \cite{Ala}.
 Moreover, the same method
has provided analytical evidence for the absence of an upper
critical dimension for the KPZ strong coupling phase \cite{Second},
 which
has been a highly debated subject \cite{debate}. Additionally, it has
also permitted to analyze the behavior of the  simpler linear
surface growth model, i.e. the
Edwards-Wilkinson (EW) equation \cite{Third}.

   In this paper we intend to go further in the application 
and understanding of
this new RSRG approach. In particular, we study
 a class of 
 systems exhibiting roughening transitions even
in one dimension, by renormalizing them with the new RSRG approach.
  Our motivation for that is twofold; on one hand
we want to test the RSRG method (which was specifically devised to deal
with KPZ growth) when generalized and applied to other physical
situations, i.e. we intend to analyze its versatility to deal
with different physical problems.
 On the other hand, by doing so we will perform a
renormalization of the class of systems exhibiting a roughening 
transition in $d=1$, that allows us
to get some insight into their interesting physics.

   The paper is structured as follows. In section I we
present the family of models exhibiting a
one-dimensional  roughening transition, and review their
main properties. In section II we present a generic
two-parametric model
 in this class suitable to be renormalized using the RSRG
approach, and discuss in detail all the different phases
and physical behaviors.
 In section III we briefly present the main traits
of the RSRG approach, discuss its application to our model and
present  a detailed discussion of the results.  In section IV the
conclusions are presented. Finally, in 
the appendix we study the connection
between the one-dimensional roughening transitions discussed
previously and directed percolation.

 \section{Systems with one-dimensional roughening transitions.}

 In this section we review the class of
models exhibiting a one-dimensional roughening transition.
The firstly studied model in this class
is the so called Polynuclear growth model analyzed
by Kerstez and Wolf (KW) \cite{KW}.  
Its dynamics  
is defined by two successive steps:
 In the first one, particles are (parallelly) 
 deposited
 with probability $p$ at each site of a one-dimensional
 lattice.
 In the second one,
the terraces (kinks) formed by the previous deposition process,
 grow laterally in a deterministic way
by $u$ units (or less if less space is available on the terrace). 
This is, kinks move deterministically increasing always the 
averaged
height \cite{KW}.
 These two processes are iterated in time.
For  every $u$ there is a critical
value of $p$, $p_c$, such that for $p > p_c$ the surface grows
homogeneously,  it is flat, and moves with maximal velocity \cite{KW}.
The roughness exponent in this flat regime is $\alpha =0$, and
$\beta = 0$ \cite{exp}.
On the other hand, for $p < p_c$ steps are less
likely to annihilate and the surface becomes rough.
In this phase the roughness exponent is estimated to be
$\alpha \approx 0.5$, and $\beta \approx 0.33$ compatible with their
 corresponding KPZ values.
The roughening transition, as we discuss below, and as first
 pointed out by KW, is related
to directed percolation (DP) \cite{DP}.

  After the seminal work by KW other models proposed
for rather different physical problems have appeared in the literature,
exhibiting similar phenomenology.

 Alon {\em et al.} \cite{AEHM} proposed a model with absorption
of particles and desorption at the edges of grown islands.
This mimics the fact that,  in crystal growth,
 particles absorbed in the interior
of grown islands
 are more strongly bounded than particles on the
edges. Their model is sequentially updated, and for large
values of the absorbing probability the system is rough
($\alpha \approx 0.5$, and $\beta \approx 0.33$, compatible with their
 corresponding KPZ values), while
for smaller values of the growing probability, the desorption mechanism
 has a larger relative importance, small islands are more easily eliminated,
and the asymptotic behavior is flat ($\alpha \approx \beta \approx 0$).
Moreover, these authors identify  also a spontaneously broken
symmetry in the model in the flat phase \cite{AEHM}.
 The phase transition in
this case is also related to DP.

More recently an apparently unrelated model has been
formulated in order to describe fungal growth. A fungal
colony grows invading an
environment, with the peculiarity that the local growth probability
is non-markovian  \cite{Lopez}. With this main ingredient, and considering a
triangular lattice, L\'opez et al. \cite{Lopez}
found a roughening transition
between a rough EW phase and a flat regime.

 All the aforementioned models share a common property:
The roughening transition can be related 
to a DP transition \cite{DP,baw}.
  Having introduced this class of DP related roughening transitions,
in the next section we present a new model, in this same 
generic class,
which turns out to be more suitable to be studied using the
RSRG.

\begin{figure}
\centerline{
\epsfxsize=2in
\epsfbox{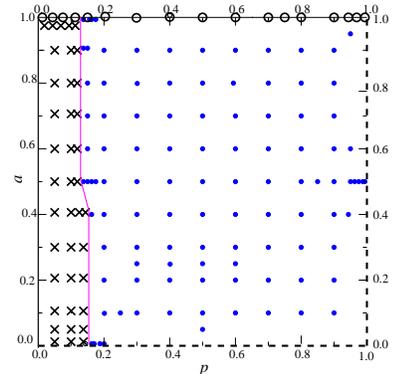}}
\caption{System phase diagram. 
Filled circles ($\bullet$)
denote points in the Edward-Wilkinson universality class;
	crosses ($\times $), points in the flat class;
empty circles $(\circ )$ points in the self-similar class
and dashed lines points in the random deposition class.
}
\label{diagr.fig}
\end{figure}

\section{The model} 

Instead of trying to directly renormalize any of the previously
 defined models, we find it more convenient to define a new model 
with the same generic phenomenology. 
The reason for this is that in the new model there is no 
proliferation of new parameters \cite{Third}
 when the RG transformation is applied,
and therefore the RG flow and the fixed point structure can be analyzed 
in a simple way. This does not exclude, in principle, the possibility of
renormalizing directly any of the previously described models.

The model describes a surface
 driven by three different physical processes:
deposition, evaporation and condensation of particles
on a one dimensional lattice, $i=1,\ldots ,L$. At
 each lattice site we associate an integer non-negative 
 variable $h(i)$. 
The random deposition process, corresponds
 to an external flux of particles,
and occurs at each time step with probability $p$. 
The other two processes, i.e. evaporation and condensation,
 decrease the
 height difference between neighbor sites, and constitute therefore
a smoothening source.
 These occur with complementary probability $1-p$ at each time step, and
they are responsible for the generation of
 correlations among different sites.

We consider flat initial configurations
($h(i)=0 \ \forall i$ at $t=0$),
 periodic boundary conditions  \cite{ob}
($h(1)=h(L) \ \forall t$), and sequential updating. 
The dynamics is defined by the following algorithm:
At each simulation step a lattice site $i_0$ is selected randomly.
Its height can either be increased by one unit with probability $p$
(random deposition)
\begin{equation}
h(i_0) \rightarrow h(i_0)+1   
\end {equation} 
or alternatively, with complementary
 probability $1-p$ the surface is
smoothened away (evaporation or condensation) in the following way; 
\begin{eqnarray}
h(i_0) & \rightarrow & \ h(i_0)+{\rm int}[a \ (h(i_0+1)-h(i_0))]  
\nonumber  \\ 
h(i_0) & \rightarrow & \ h(i_0)+{\rm int}[a \ (h(i_0-1)-h(i_0))]   
\label{1aprile} 
\end{eqnarray}
 where $a \in [0,1]$.
Each one of these two possible events occurs with equal
 probability $(1-p)/2$. 
This process causes a decrease of the height difference (smoothening)
 between the site $i_0$ and one of its neighbors.
In the dynamical rule  given by Eq. (\ref{1aprile})
we have introduced the integer part function to
enforce the height variables to take integer values. 

Let us now discuss the model phenomenology.
We have investigated the parameter space $(p,a)$ with 
 analytical and computational methods in different points
 as shown in Fig.$\ref{diagr.fig}$. 
\begin{figure}
\centerline{
\epsfxsize=3.5in
\epsfbox{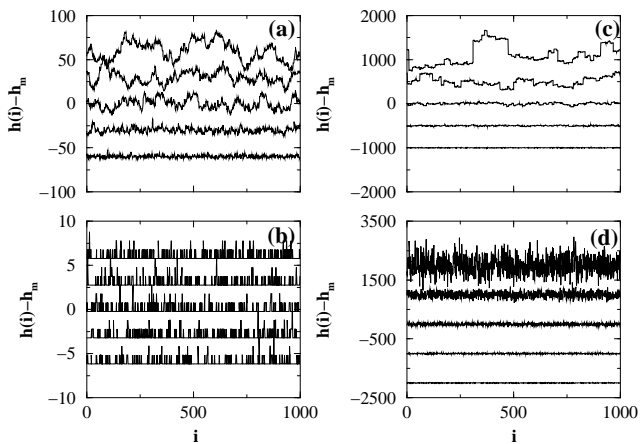}}
\caption{Evolution profiles
 for the different universality class  at times  (from bottom 
to top) $t=10,
 10^2,
10^3, 10^4 $ and $10^5$ for $L=1000$.
 (a) $(p=0.4,a=0.5)$ corresponding to the
 Edwards-Wilkinson class
 (b) $(p=0.1,a=0.5)$ in the flat class;
 (c) $(p=0.75,a=1.0)$ in the self-similar class; and
 (d) $(p=1, a=0.5) $ random deposition class.
 Curves at $t=10^4$ and $10^5$ ($t=10$, $10^2$)
 are shifted upward (downward)
 for the sake of clarity.}
\label{outlines.fig}
\end{figure}
The evolution of the surface width
can belong to one out of four different
scaling regimes
 depending on the parameter values, $p$ and $a$. 
These four universality classes define
 four qualitatively different  growth morphologies, the
corresponding typical 
  surface profiles of which
 are shown in Fig. $\ref{outlines.fig}$. 
They correspond to the following universality classes:
 The {\it Edwards-Wilkinson} \cite{EW,HZ,Krug,Laszlo},
the {\it flat},
 the {\it self-similar},  and
the {\it random deposition} \cite{HZ,Krug,Laszlo} universality class.
\begin{figure} 
\centerline{ 
\epsfxsize=6in 
\epsfysize=7in
\epsfbox{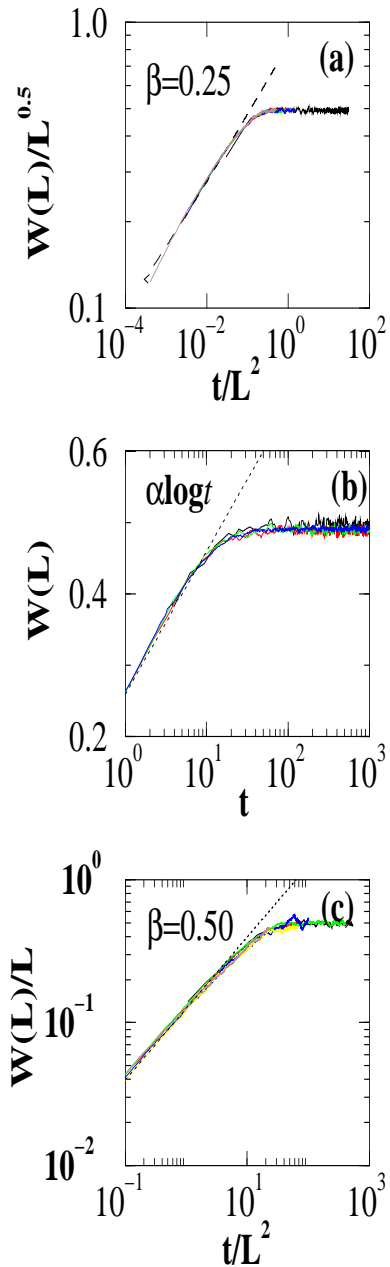}} 
\caption{
Width evolution for system sizes
 $L=32, 50, 64, 90, 128$, and $256$,
 for three different pairs of parameter values corresponding 
to: (a) the Edwards-Wilkinson universality class, $(p,a)=(0.4,0.5)$; 
(b) the flat universality class, $(p,a)=(0.1,0.5)$; and
 (c) the self-similar universality class $(p,a)=(0.75,1.0)$. 
In (a) and (c) the width is rescaled with $L^{\alpha}$  
 and plotted
 versus the time rescaled with $L^z$, with
$\alpha=0.50\pm 0.01$, $\beta=0.25\pm 0.01$  in  (a),
and  $\alpha=1.00 \pm 0.02$, $\beta=0.50\pm 0.03$ in (c). 
In (b)
  a semilogarithmic scale is used;
 there is a logarithmic dependence on time 
 for 
 $t<10$, and a system size independent saturation width, compatible with
 $\alpha=0.00\pm 0.01 $ and $\beta\sim 0$.}     
\label{width.fig} 
\end{figure}

\subsection{Edwards-Wilkinson universality class} 

The EW class is
 characterized by critical
 indexes $\alpha=1/2$, $\beta=1/4$ and $z=2$ \cite{HZ,Krug,Laszlo}.
We observe this type of scaling
 for parameter values
 $p>p_c(a)$, (where $p= p_c(a)$ is the separatrix
 between this phase and the flat one)
 with $p\ne 1$ and $a \ne 0, 1$
(see Fig. \ref{diagr.fig} where points
 in this class are marked with filled circles). 
In Fig. \ref{width.fig}
 (a) we show the collapse of the surface width
for different system sizes
(see figure caption).
The best curve collapse is obtained for
 $\alpha =0.50\pm 0.01$ and $\beta=0.25\pm 0.01$ in agreement
with the system being in the Edwards-Wilkinson universality class.
This phase has a mean growing velocity
different from zero (i.e. grows continuously in time)
and has a rough surface appearance
as shown in Fig. \ref{outlines.fig}(a).

Let us present here a simple argument showing why this rough
phase is EW like and not KPZ like.
For that, we calculate
the mean over different runs of the
quantity $h(i,t+\Delta)-h(i,t)$ (where $\Delta t=1/L$ is a time increment).
We can write: 
\begin{eqnarray}
 && \langle h(i,t+\Delta t)-h(i,t) \rangle 
  = \nonumber \\ 
& & {\frac{p}{L}}
 +{\frac{1}{L}\ \frac{(1-p)}{2}}
 \ a \ \left( h(i+1,t)-h(i,t)\right)+ \nonumber \\ 
   \\ 
& & {\frac{1}{L}\
 \frac{(1-p)}{2}} \ a \ \left(h(i-1,t)-h(i,t)\right) \nonumber \\  
    \\ 
& & {\frac{p}{L}} +
 { \frac{(1-p)}{2 L}}
 \ a\ \left(h(i+1,t)+h(i-1,t)-2h(i,t)\right) 
\label{equ}
\end{eqnarray} 
where it can be seen that there are two different contributions 
to the local velocity at each site $i$; the contribution 
coming from deposition, proportional to $p$, and 
the term arising from the smooth away processes. Observe that in 
this calculation we have neglected 
the effect of the integer part function,
which we assume to be irrelevant
in this phase (i.e. we expect it to reduce the mean velocity
but not to change qualitatively the behavior).

Dividing Eq. (\ref{equ}) by $\Delta t$
and performing the thermodynamic limit, i.e.,
 $L \rightarrow \infty$ (or equivalently $\Delta t \rightarrow 0$),
 we obtain: 
\be 
\frac{\partial h}{\partial t}=p+ \frac{(1-p)}{2} \ a \ {\nabla}^2 h. 
\label{equ2} 
\ee 
that is the Edwards-Wilkinson
equation driven by an external force $p$.
For $p=1$ or $a=0$
this equation describes random deposition as, in fact, is the
case in our model.
Eq. (\ref{equ2}) fails to describe
 the growth of our model in the cases $p\le p_c$ or $a=1$. 
In fact, as stated before, in Eq. ($\ref{equ}$)
 the integer part function has been neglected and this approximation
is incorrect
 when: (i) $p \le p_c$, the fluctuations of the surface are small,
and the rounding off mechanism due to the integer part function
takes over, pinning and flattening the surface,
 and
(ii) in the case $a=1$, for which, 
in the absence of smoothening, the continuum limit has to be taken
more carefully. In this last case, we expect the dynamics to be
controlled by the diffusion of height steps.
 
\subsection{Flat universality class} 

This universality class
is characterized by critical indexes
$\alpha=0$ and $\beta=0$ (with logarithmic corrections).
We observe this phase for parameter values:
$p < p_c(a)$ with $p \neq 0$ and $a\ne 0,1$.
In Fig. $\ref{diagr.fig}$ we plot points
 in this class with crosses in the parameter space.
In this phase height fluctuations are
 independent on sample size (i.e. $\alpha=0$), 
the mean surface velocity
is zero in the thermodynamic limit, 
  and the lowest level $(h=0)$
is occupied with a finite density
 as shown in Fig. \ref{outlines.fig}(b).
 Due to finite size effects,
finite systems in this class may have a non-vanishing velocity.
                                                            
In Fig. \ref{width.fig}(b) we have plotted in a semilogarithmic scale
 the surface width versus time for different system sizes
(see figure caption).
Observe that the width has a logarithmic dependence on
 $t$ for short times 
\begin{equation} 
W(t)\propto \log(t). 
\end{equation} 
For all the different points in this phase, by employing data
collapse techniques we evaluate  
$\alpha=0.00 \pm 0.05$ and $\beta = 0$ with logarithmic 
corrections.

The existence of this phase is due to the fact
that for small values of $p$ the deposition
process is much unlikely than the smoothening one, and there
is a physical constraint preventing the surface to go below
the lowest, $h=0$ level.
The integer part function in the dynamic rules favors the
evaporation process by giving and extra negative drift
term with respect to Eq. \ref{equ2}. This extra term binds 
the surface to the lowest level making it flat for small values of $p$.
In particular, 
for a fixed value of $a$, the roughening transition corresponds to the 
value of $p$ 
for which this extra negative term is equal to $ - p$. For values 
of $p$ slightly larger than that critical value the surface
unbinds and grows with constant velocity.

\subsection{Self-similar universality class} 

This class
is characterized by
$\alpha=1$, $\beta=1/2$ and $z=2$, it is observed 
for parameter values $a=1$ and $p\ne 0,1$ which are 
plotted in
Fig. \ref{diagr.fig} with empty circles.
In Fig. \ref{width.fig}(a) we show the collapse of the width
for different system sizes (see figure caption);
from the best data collapse we measure
$\alpha =1.00\pm 0.02$ and $\beta=0.50\pm 0.03$.
The surface in this regime is therefore self-similar, that is,
height fluctuations are of the order of magnitude of the system size
(see also Fig. \ref{outlines.fig}(c) where a typical profile is
shown). In this case, contrarily to $a \neq 1$ we observe no roughening 
 transition (except for a trivial one at $p=0$.)

Observe that for $a=1$ the dynamical rules can be written as: 
\be 
\begin{array}{lll} 
h(i_0\ ,t)=h(i_0\ ,t-1)+1 & \mbox{with prob.} & \ p \\ 
 &  \\ 
h(i_0\ ,t)=h(i_0+1\ ,t-1) & \mbox{with prob.} & \ (1-p)/2  \\ 
&  \\ 
h(i_0\ ,t)=h(i_0-1\ ,t-1) & \mbox{with prob.} & \ (1-p)/2, 
\end{array} 
\label{diffu} 
\ee  
where the integer part function, being redundant, has been
omitted. Therefore, the extra negative driving term, arising 
as a consequence of the integer part function in the dynamics
(which, as discussed in the previous subsection
favors evaporation) does not exist in this case. This explains 
why there is  
no roughening transition for $a=1$.
 
   The main qualitative physical difference between this phase
and the EW and the flat phases is that here there is no 
smoothening of height gradients. Large steps diffuse
in space
due to the effect of the second and third rules in Eq.  (\ref{diffu}),
but do not smoothen away, creating a much rougher surface.

 Now we study 
the mean field solution of this case, obtained by neglecting
spatial correlations. This turns out to be an useful 
calculation as we will show afterwards.

The master equation for the probability
 $P(h,t)$ for a given site to have a height
 value $h$ at time $t$ is
\bea 
\begin{array}{lcl} 
\displaystyle{\frac{\partial
 P(h,t)}{\partial t}} & = & pP(h-1,t)-pP(h,t)+ \\ 
 & & \\ 
 & + & (1-p) P(h,t)- (1-p)P(h,t) \\ 
 & & \\ 
 & =& p\left[P(h-1,t)-P(h,t)\right] 
\end{array} 
\label{Pmaster} 
\eea 
where we have assumed nearest neighbor sites
 to have the same probability
distribution as the site under consideration.
Eq. (\ref{Pmaster}) is the same equation
that we would obtain if we considered a mean field approximation of the 
random deposition process. 
In order to solve (\ref{Pmaster}), it is convenient to 
introduce the generating function defined as
\be 
G(x,t)=\sum_{h=0}^{\infty} {x^h P(h,t)} 
\label{Gdef} 
\ee 
with the normalization condition $G(1,t)=1\ \forall t$. 
The master equation written in terms of $G(x,t)$ reads 
\be 
\frac{\partial G(x)}{\partial t}=p\left[ xG(x,t)-
G(x,t)\right]=p(x-1)G(x,t).
\label{Gt} 
\ee 
By taking derivatives with respect to the dummy variable $x$,
 one can write
\bea 
\begin{array}{rcl} 
W^2(t) & = & {<h^2-{<h>}^2>}= \\ 
& = & \left.\left[\partial^2_x G(x)+
 \partial_xG(x)\right] \right|_{x=1}-
{\left[\left.\partial_xG(x)\right|_{x=1}\right]}^2.
\end{array} 
\label{senzanome}
\eea 
Taking time derivatives of both sides
 and integrating the resulting equation,
we finally obtain
$W(t) =\sqrt{pt}$, and consequently $\beta=1/2$. 
On the other hand, the height differences 
among neighbors (or steps) perform 
a random diffusion in the direction perpendicular to the growth. 
Like in the case of a random walker
the average step displacement scales as square root of time.
But the effective time for step displacement is $(1-p) t$,
and therefore in order to cover a distance $L$, a characteristic time     
\bea 
t \sim {L^2}/{(1-p)}
\label{tL.eq} 
\eea 
is needed.
Consequently the dynamical exponent is $z=2$. 
Combining this result with the fact that $\beta=1/2$ we obtain 
 $\alpha=1$ just by using the scaling
 relation $\alpha=z/\beta$.
\begin{figure} 
\centerline{ 
\epsfxsize=3in 
\epsfbox{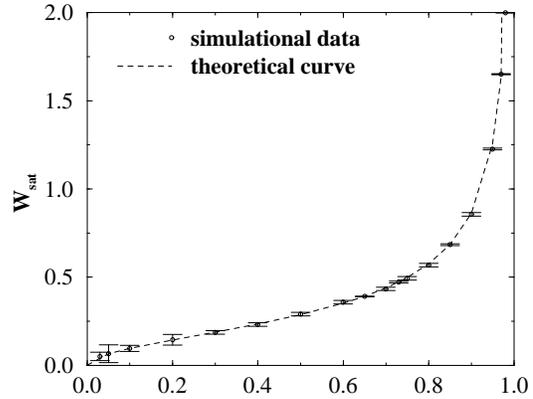}} 
\caption{Saturated width of a system of
 size $L=64$ with $a=0.5$, as a function of $p$. 
The data are extremely well fitted by the mean field prediction
}    
\label{wp.fig} 
\end{figure} 
Therefore, the values of the critical exponents
found in mean field approximation
agree perfectly with the numerical results.
Moreover, the mean field approximation also reproduces
the dependence
of the saturation width as a function of $p$
for a fixed system size. 
In fact, using $W(t)=\sqrt{pt}$  and Eq. (\ref{tL.eq})
we find: 
\bea 
W_{sat}(p)\sim\sqrt{\frac{p}{(1-p)}}  
\label{mf.eq} 
\eea 
in perfect agreement width the simulation
results as shown in Fig. \ref{wp.fig}. 

\subsection{Random deposition universality class}

The random deposition class is
characterized by $\alpha= \infty$ and $\beta=1/2$.
With dashed lines in Fig. $\ref{diagr.fig}$ we have plotted
the two lines in this class, namely those given by the conditions
 $a=0$ and $p=1$ respectively.
For these values
the only physical mechanism present 
is random deposition of particles.
In this phase the surface is spatially uncorrelated, and
its roughness increases rapidly with time,
with $\beta =1/2$, and there is no width saturation ($z=\infty$)
as shown in Fig. \ref{outlines.fig}(d).

\subsection{The roughening transition} 
 
As stated previously, both in our 
model and in the previously discussed ones 
the transition between the flat and the rough
phase belongs to the directed percolation
 (DP) universality class. 
The reason for this can be explained by mapping
the dynamics in a directed percolation like model. 
In order to do this in our model (for the others similar mappings
can be performed), we introduce a new set of
 variables $\{s_i\}$ defined  as 
\bea 
\begin{array}{lcr} 
{\it s_i}=1 & {\rm if} & h(i)=0 \\ 
{\it s_i}=0 & {\rm if} & h(i) \ne 0. 
\end {array} 
\eea 
By studying
the evolution of these variables 
we focus our attention on the lowest height level occupation. 
 This information is enough to describe the phase transition \cite{KW},
as we show in what follows.
 A site $i$ with $s_i=1$ corresponds to an occupied (active) site 
in a DP like model (or better, in a contact process model
which is a sequentially updated version of DP \cite{DP}).
 On the contrary,
$s_i=0$ for empty (absorbing) sites of the DP like model. The 
deposition process may change with probability $p$ an occupied, 
$s_i=1$, site
into an unoccupied, $s_i=0$, site. This same mechanism is also present 
in the contact process \cite{DP}. 
On the other hand, the smoothening mechanism,
occurring with probability $1-p$,
may induce \cite{note} the ''infection'' of an empty site by a neighboring 
active site as in the contact process. 
In this language,
the flat surface phase corresponds
to the DP active phase, and the rough phase to the absorbing one.
The key feature is that in absorbing regions (this is, in regions
with $s_i=0$), activity cannot be generated spontaneously. This 
is the main physics of DP and therefore the critical behavior at
the roughening transition is related to the transition into an 
absorbing phase 
of DP and related models. 

By using this mapping the scaling of some magnitudes 
can be related to DP exponents.
 In particular, the density of sites
at the lowest level, $n_0$, should scale as a function of 
the distance to the critical point, $\varepsilon=| p-p_c |$, like
$n_0 \sim \varepsilon^{\beta_{DP}}$. Right at the critical point
$n_0$ decays in time as $n_0(t) \sim t^{-\theta_{DP}}$. For finite
systems, of size $L$, 
\be
n_0 \sim L^{-{\beta_{DP}}/{\nu_{\bot,DP}}} \sim L^{-x_f}
\label{xf}
\ee
where
$\nu_{\bot,DP}$ is the correlation length exponent.
Finally, the mean surface velocity 
is inversely proportional to the life time
$\tau={
|\varepsilon|}^{-\nu_{\|,DP}}$ of the DP active phase, where
 $\nu_{\parallel,DP}$
is the usual correlation time exponent.
Therefore,
\be
v \sim  {\varepsilon}^{-\nu _{\|}}
\sim L^{{-\nu_{\|}}/{\nu_{\bot}}} \sim L^{-x_v}. 
\ee
The one-dimensional DP values of the previously introduced 
exponents are:
$\beta_{DP}=0.27649(4)$, $\theta=0.15947(3)$, $x_f=0.25208(5)$
and $x_v=1.58074(4)$ (numbers in parentheses denote uncertainties
in the last figure) \cite{rr}.

In order to verify numerically if the previous prediction holds we have
studied extensively the $a=0.5$ case. 
We have performed numerical simulations in order
to determine critical exponents, to be compared with their
corresponding DP values.
In particular, by considering a system size  $L=3000$, and
averaging over $8000$ independent runs, we measure
$n_0$ as a function of time.
A power law decay is observed at (the critical point)  
$p_c=0.13740 \pm 0.00005$, with an associated exponent
\be
\theta=0.160 \pm 0.003
\ee
in perfect agreement with its
DP value.
\begin{figure} 
\centerline{ 
\epsfxsize=3.5in 
\epsfysize=5in
\epsfbox{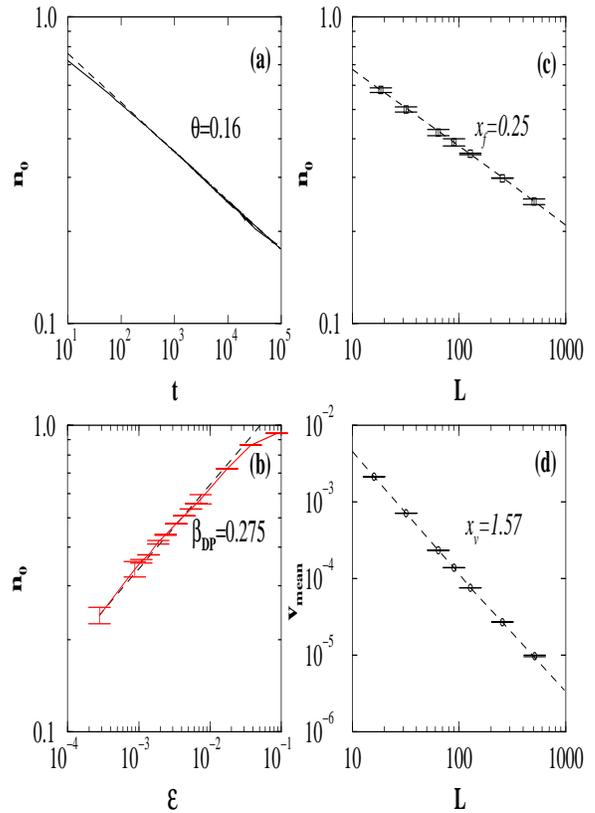}} 
\caption{Critical indexes characterizing the DP transition. 
(a)  Log-log plot of the 
density of sites at lowest level as a function of 
time  at 
  $(0.1374, 0.5)$;
the associated critical exponent is  $\theta=0.160 \pm 0.003$. 
 (b) Log-log plot of the
asymptotic density of states at the lowest site
 versus $\varepsilon=p_c-p$;
 from the slope we measure $\beta_{DP}=0.275 \pm 0.007$. 
In  (c) and (d) 
 we  report the density of
 sites at the lowest level and the mean velocity, at the
 critical point  for different system sizes.
The corresponding critical indexes are 
 $x_f=0.250 \pm 0.003$ and $x_v=1.57 \pm 0.01$.} 
\label{DP.fig} 
\end{figure} 
Using the previously estimated 
value of the critical point, we measure
 the stationary density of sites at the lowest layer
 $h=0$ for $p<p_c$. 
In Fig. $\ref{DP.fig}$(b)
 we show the data for $\varepsilon \in [{10}^{-4},{10}^{-1}]$. 
Performing a fit  for small values of $\varepsilon$ we find
\be 
\beta_{DP}=0.275 \pm 0.007 
\ee         
also in good agreement with its DP value. 

    By performing a finite size scaling analysis, we determine 
$x_f=0.250\pm 0.03$ and $x_v=1.57\pm 0.01$ to be compared 
with their corresponding DP values $x_f=0.25208(3)$ and $x_v=1.5807(5)$
respectively. Numerical data are showed in Fig. \ref{DP.fig} (c)-(d).

  Summing up, all the measured exponents confirm the hypothesis
that the roughening transition is controlled by a DP fixed point.
 
\section{Renormalization Approach} 
 
Having described in detail the model phenomenology, in
this section we apply 
the RSRG technique
recently proposed \cite{First,Second,Third}
to renormalize our model. 
  For a detailed description of the method we refer
the reader to reference \cite{Third}. In a nutshell
the RG method consist of two main ingredients.
 One 
is the definition of cells of generic substrate length $L_k=2^kL_0 $
(where $L_0$ is the length of the minimal relevant substrate scale)
 and 
height $h_k$, with which the space in which the surface grows
 can be covered; cells are progressively invaded by the
growing surface.
 The second
one is the identification of the effective dynamics of these cells 
at a generic scale;
this is, one studies the effective rules by which
 cells are progressively 
invaded by the growing surface at a given scale.
This effective dynamics is described by parameters that change
upon changing the scale of description. In our case we have
a set of parameters $ {\bf x}_k =(a_k, p_k)$ at scale $L_k$. 
Known the effective dynamics at scale $L_k$,
defined by ${\bf x}_k$ and the associated width 
at that scale $W_{k}$,
the renormalization problem consists in determining  
what are the width values $W_{k+ {\it m}}$
 and the effective dynamic parameters
 $ {\bf x}_{k+ {\it m}}$ at a coarser scale, $L_{k+ {\it m}}$.    
This problem was analyzed in  \cite{First,Second,Third}, with 
the final conclusion that one can write 
\be 
W^2_{k+{\it m}}=W^2_k F_{\it m}({\bf x}_k) 
\ee 
 where $F_{\it m}({\bf x}_k)=1+4 w^2({\it m},{\bf x}_k)$ and 
$w({\it m},{\bf x}_k)$ 
is the width of a system composed of
 ${\it 2^m}$ cells of unitary height
 with a dynamics driven by the parameters ${\bf x}_k$.   
This last functions can be determined by means of a rather
inexpensive MonteCarlo simulation. 
\begin{figure} 
\centerline{ 
\epsfxsize=3.5in 
\epsfbox{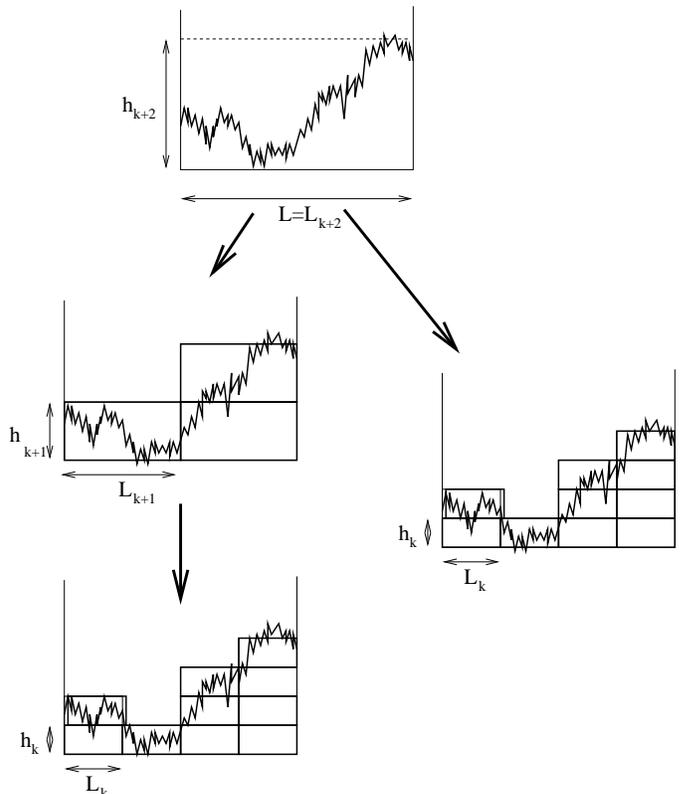}} 
\caption{A system of size 
$L_{k+2}=4L_k$ can be 
studied as composed of 
four cell of size $L_k$, or alternatively by iterating 
twice a partition in cells of half length. See text. 
} 
\label{rin.fig} 
\end{figure} 
The width at scale $k+2$ can be calculated in two different ways; 
(i) by direct use of the previous formula with $m=2$, or
alternatively, (ii) by iterating twice the previous transformation
with $m=1$. Imposing that both the previous procedures give 
the same result for the width $W_{k+2}$, 
one obtains a renormalization condition, namely,
\cite{First,Third}
\be 
x_{k+1} = F_1^{-1}[F_2(x_k)/F_1(x_k)]. 
\label{pri}
\ee 
For mono-parametric models this equation is enough to determine 
the evolution under RG transformations of the effective 
parameter $x_k$, and from it all the scale invariant physics
can be elucidated \cite{First,Second,Third}.
In the present two-parametric dynamics 
it is necessary to consider another independent analogous 
equation, corresponding to calculating the width at scale $k+3$
in two different ways; 
\bea
\left\{ 
\begin{array}{l} 
{\bf x}_{k+1}= F_1^{-1}[F_2({\bf x}_k)/F_1({\bf x}_k)] \\
{\bf x}_{k+1}= F_2^{-1}[F_3({\bf x}_k)/F_1({\bf x}_k)].   
\end{array} 
\right.  
\label{flux.eq} 
\eea
The first equation is just Eq. (\ref{pri}), and the second 
 states that the width obtained by dividing the system in 8 cells,
should be the same as the width obtained dividing
first the system in two blocks, and then each of these blocks
in four sub-cells. 
 This set of equations give the (discrete) flux of the 
effective dynamics parameters upon coarse graining. 
The scale invariant dynamics  is determined by the 
fixed point (or points) of Eq. (\ref{flux.eq}),
 ${\bf x}^*$ for which
\bea
\left\{
\begin{array}{l}
F_1({\bf x}^*)=F_2({\bf x}^*)/F_1({\bf x}^*) \\
F_2({\bf x}^*)=F_3({\bf x}^*)/F_1({\bf x}^*)  
\end{array}
\right.
\label{fissi.eq}
\eea
 is satisfied.
Once ${\bf x}^*$ is known
 the $\alpha$ exponent is easily determined by 
\be
\alpha=\frac{1}{2} \log_2(F_1({\bf x}^*)).
\label{alfa}
\ee
 
In order to evaluate the
 $F_1$, $F_2$ and $F_3$
 functions we have performed
 MonteCarlo simulations 
 for  system of size $L=2,4,8$, and averaged, once the stationary 
state is reached, for long enough times.
We have determined these functions
on the sites of a  $100 \times 100$ lattice in the parameter
space, this is,
for values of $p$ and $a$ multiples of $0.01$. 
 
\begin{figure} 
\centerline{ 
\epsfxsize=3in 
\epsfbox{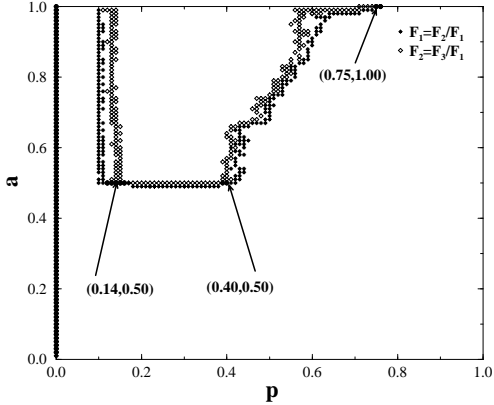}} 
\caption{ 
The two lines (marked respectively with diamonds and empty circles)
correspond to the values for which
$ F_1({\bf x})={F_2({\bf x})}/{F_1({\bf x})}$
and $F_2({\bf x})={F_3({\bf x})}/{F_1({\bf x})}$, within the
numerical error respectively.
The intersection points of these
two lines define  the dynamic fixed points, invariant under RG flow.
See text.
} 
\label{fissi.fig} 
\end{figure}

In Fig. \ref{fissi.fig} we show the 
curves corresponding to
 $F_1=F_2/F_1$ and $F_2=F_3/F_1$. 
 The intersection points of these curves are the RG fixed points
 within numerical accuracy.
We find the following stable fixed points: 
\bea 
\begin{array}{llr} 
p=0.75, & a=1 & {\rm corresponding \ to}\  \alpha=1.00\pm 0.02\\ 
p=0.40, & a=0.5 & {\rm corresponding \ to} \  \alpha=0.50 \pm 0.01 \\ 
p=0, & \forall a & {\rm corresponding \ to} \  \alpha=0.00 \pm 0.01 
\end {array} 
\eea 
while at $(p,a)=(0.14,0.5)$ there is an unstable fixed point.  
The stability is determined by observing whether nearby points
flow into the fixed point or flow away from it, under application
of Eq. (\ref{flux.eq}), and the corresponding values of
$\alpha$ are determined using Eq. (\ref{alfa}).
 
\begin{figure}
\centerline{
\epsfxsize=3in
\epsfbox{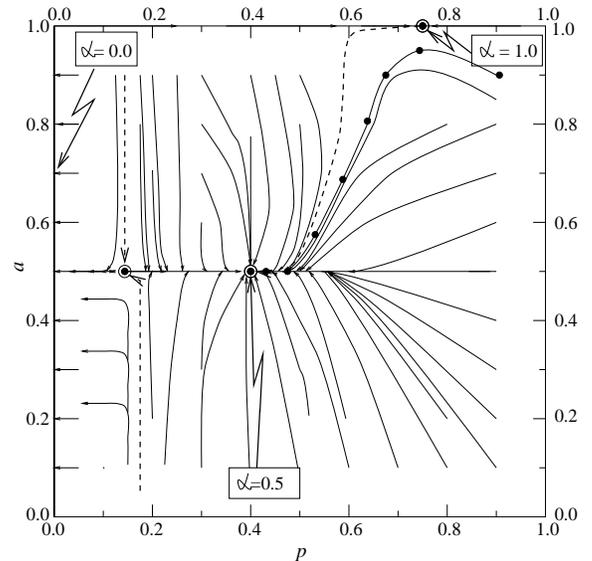}}
\caption{Renormalization Group flow
 for Eq. ($\ref{1aprile}$).
 The stable fixed points are
 $p=0.4, a=0.5$ (EW behavior);
 $p=0.75, a=1$ (self-similar phase);
 $p, \forall a$ (flat phase);
 and the unstable point $p=0.14, a=0.5$
 (corresponding to a DP transition).
 For the flux line starting at $(0.95,0.95)$,
 we show explicitly the points
 found by iterating the renormalization group
 equations.
 To guide the eye 
discrete sequences of points obtained by iteration of the RG 
transformation are plotted as continuous lines.}
\label{flusso.fig}
\end{figure}

We have also determined different
renormalization flux lines as shown in Fig. \ref{flusso.fig}. 
Observe that each continuous flow line in 
 Fig. \ref{flusso.fig}
 has been constructed by  joining together
 consecutive points obtained
 from the renormalization group equations \ref{flux.eq} (which 
determines a discrete iterative
 mapping and not a continuous flow). Observe
that the lines defined by $a=0.5$
 and $a=1$ define invariant manifolds.

The accordance between
the renormalization
flow diagram (Fig. \ref{flusso.fig})
and the phase diagram independently
found with simulation on large systems (Fig. \ref{diagr.fig})
is very good (compare Fig. \ref{diagr.fig} and
\ref{flusso.fig}).  In particular:
\begin{itemize}
\item  For all points yielding in the EW phase we find trajectories 
converging to the stable fixed point $(0.4, 0.5)$ with
$\alpha=0.5 \pm 0.01$, in perfect agreement with the EW value.
\item The separatrix in Fig. $\ref{diagr.fig}$ is reproduced 
quite accurately in Fig. $\ref{flusso.fig}$, in particular,
the unstable 
fixed point $(0.14,0.5)$ is located on it. 
\item Points to the left of the separatrix (i.e. points in the
flat phase) 
flow towards the fixed line at $p=0$ for which $\alpha=0$ 
(as corresponds to the flat phase).
\item  All points with $a=1$ and $p > 0$, i.e. in the self-similar
phase,  flow to a fixed point
at $(0.75,1)$ with $\alpha=1.00 \pm 0.01$. 
\end{itemize}
                                  
 We have verified all the above conclusions to be stable when 
more refined RSRG algorithms are considered. This is, instead 
of considering by-partitions and quadri-partitions of a given
growing surface, 
one can consider larger partitions, and the technique 
described above remains the same in spirit \cite{First,Third}.
By doing
this, we observe all the above described 
results to remain unaltered. This stability upon changes in the
RSRG details supports 
the fact that no new parameter has to be introduced upon renormalization 
to describe he scale invariant dynamics \cite{Third}, i.e. 
the scale invariant dynamics is well described at an arbitrary 
scale by the found fixed points.

\section{Conclusions}

We have discussed different growth models exhibiting a
one-dimensional roughening transition. In particular, we
have introduced
a new two-parametric model capturing the main physics 
of this type of roughening transition, i.e exhibiting
a rough and a flat phase, and a critical line separating
them, related to directed percolation. The new model 
has the technical advantage that no new parameter 
proliferate upon renormalization. Apart from the
previously discussed phenomenology, the new model 
exhibits also a self-similar phase. All the different 
 phases and transitions have been analyzed by means
of extensive numerical simulations and some analytical 
approaches. 

We have renormalized this model by using a 
recently introduced RSRG approach. In particular,
fixed points, corresponding to the scale invariant dynamics are found,
 and the corresponding roughness exponent determined.
The results are in perfect accordance with all the 
numerical and analytical findings. In particular, the
phase diagram is perfectly reproduced: The rough, flat, and
self-similar phases, as well as the separatrix among them, 
and their associated roughness exponents
are identified with great accuracy.

  This confirms the general validity of the RSRG method
to deal with anisotropic fractal growth
in cases others than KPZ growth for which
it was explicitly design.

We acknowledge interesting discussions  
 with 
M. Marsili, G. Parisi, Y. Tu, and W. Genovese.
 We thank specially C. Castellano 
for very useful comments, suggestions and a critical 
reading of the manuscript.
This work has been partially  
supported by the European network contract FMRXCT980183.


\begin{thebibliography}{99}
\bibitem{exp}
The usual scaling ansatz for the width $W$, is
$W(L,t)={<{(h-<h>)}^2>}^{1/2}$
in a substrate of length $L$
at time $t$ is
$W(L,t)=L^{\alpha}f(t/L^z)$
with $f(x)\sim x^{\beta} \ {\rm for}\ {x\ll 1}$,
and $f(x)\sim {\rm cost.} \ {\rm for}\ {x\gg 1}$.
The critical indices $\alpha$,
$\beta$ and $z$ are  connected by $z=\alpha/\beta$.
\bibitem{ob} We have verified that all our conclusions remains
unchanged when open, instead of periodic,  boundary conditions
are considered.
\bibitem{note} Observe that in the case $a=1$, given the absence of
the integer part function this infection mechanism
is not present.
\end{thebibliography}
\end{document}